\documentclass{article}
\usepackage{graphicx}  
\usepackage{amsmath}   
\usepackage{amssymb}   
\usepackage{bm} 
\usepackage{dcolumn}
\usepackage{color}
\usepackage{mathrsfs}
\usepackage{amsfonts}
\usepackage{varioref}
\usepackage[utf8]{inputenc}
\usepackage{amsmath}
\usepackage{amsfonts}
\usepackage{amssymb}
\usepackage{enumitem}
\usepackage{float}
\usepackage{epsfig}
\usepackage{amsfonts}
\usepackage{latexsym}
\usepackage{hyperref}
\usepackage{setspace}
\usepackage{bm}
\usepackage{slashed}
\usepackage{graphicx}
\usepackage{bm}
\usepackage{hyperref}
\usepackage{dsfont}
\usepackage{caption}
\usepackage{subcaption}
\def\be{\begin{equation}}
\def\ee{\end{equation}}
\labelformat{section}{Section #1} 
\labelformat{subsection}{Section #1} 
\labelformat{subsubsection}{Section #1}
\labelformat{subsubsubsection}{Section #1}
\labelformat{equation}{Eq.~(#1)} 
\labelformat{figure}{Fig.~#1} 
\labelformat{subfigure}{Fig.~\thefigure#1} 
\labelformat{table}{Tab.~#1} 
\labelformat{appendix}{Appendix #1}
\title{\bf Dirac fermion, cosmological event horizons   and quantum entanglement }
\author{Sourav Bhattacharya\footnote{sbhatta@iitrpr.ac.in}, ~~Shankhadeep Chakrabortty\footnote{s.chakrabortty@iitrpr.ac.in} ~~and ~Shivang Goyal\footnote{2017phz0003@iitrpr.ac.in}\\
{{\small  Department of Physics, Indian Institute of Technology Ropar}}\\ {{\small Rupnagar, Punjab 140 001,
India}}}
\begin{document}
\maketitle
\begin{abstract}
\noindent We discuss the field quantisation of a free massive Dirac fermion  in the two causally disconnected static patches of the de Sitter spacetime, by using mode functions that are normalisable on the cosmological event horizon.  Using this, we compute the entanglement entropy of the vacuum state corresponding to these two  regions, for a given fermionic  mode. Further extensions of this result to more general static spherically symmetric and stationary axisymmetric spacetimes are discussed.  For the stationary axisymmetric Kerr-de Sitter spacetime in particular,  the variations of the entanglement entropy with respect to various eigenvalues and spacetime parameters are depicted numerically. We also comment on such variations when instead we consider the non-extremal black hole event horizon of the same spacetime.
\end{abstract}
\vskip .2cm
\noindent
{\bf keywords :} de Sitter, cosmological horizon, fermionic entanglement, stationary axisymmetric  spacetimes
\bigskip
\section{Introduction}
The de Sitter spacetime is the simplest solution of the Einstein equation with a positive cosmological constant, $\Lambda$. It is maximally symmetric with isometry group $SO(4,1)$ in dimension four. Its physical significance is chiefly twofold. First, owing to the observed accelerated expansion of the current universe, there seems to be a strong possibility that our current universe is endowed with a small but positive $\Lambda$, or some alternative form of the dark energy. Second, the observed high degree of spatial homogeneity and isotropy in all directions in the sky at large scales indicates that the very early universe might also have went through a phase of a  rapid accelerated expansion, known as the inflation, see, e.g.~\cite{Weinberg:2008zzc} and references therein. 

The positive $\Lambda$ is the simplest and phenomenologically a very successful  model of the dark energy.   Even though the exact nature/form of the dark energy is yet far from being well understood, it is reasonable  to expect that the de Sitter spacetime would  qualitatively model at least some salient features of any cosmological spacetime undergoing  accelerated expansion. In particular, the highest degree of symmetry present in this spacetime makes many computations doable analytically.  

There has been a tremendous effort  over decades to explore various aspects of quantum fields living in a de Sitter universe. A complete review on this topic is far from the scope of this paper. We refer our reader to~\cite{Bunch:1978yq, Mottola:1984ar, Allen:1985ux} for various aspects 
of particle creation and vacuum states in the cosmological de Sitter spacetime. See~\cite{Frob:2014zka} and references therein for a study on the Schwinger effect in de Sitter. See e.g.~\cite{Gibbons:1977mu, Otchik:1985ih, Higuchi:1986ww, Bousso:1997wi, Nojiri:2013su, Qiu:2019qgp} for aspects of particle creation and thermal effects in the static de Sitter or de Sitter black hole spacetimes. Further, we refer our reader to e.g.~\cite{Woodard:2014jba, Miao:2018bol, Moreau:2018lmz} (also references therein) for discussions on the late time non-perturbative infrared effects in the cosmological de Sitter spacetime. See also~\cite{Benisty:2019jqz, Benisty:2019pxb} for discussions on fermion driven inflation.

A natural and interesting aspect of the de Sitter space is the relativistic quantum entanglement of fields, which is the focus of this paper. If we consider an `in' vacuum state in the cosmological de Sitter spacetime, due to the accelerated expansion, the state may evolve in the future to a different or `out' vacuum state, indicating particle pair production. Such pairs turn out to be entangled. On the other hand, due to the accelerated expansion, all parts of the de Sitter space cannot be causally connected. Quantum fields living in various causally disconnected parts of de Sitter can show very non-trivial aspect of quantum entanglement. We refer our reader to~\cite{Maldacena:2012xp}-\cite{Narain:2018wlz} and references therein for a study of quantum field theoretic entanglement in the cosmological and hyperbolic coordinatisation of de Sitter. We further refer our reader to~\cite{Dong:2018cuv, Arias:2019pzy, Geng:2019bnn, Geng:2019ruz} and references therein for a study of holographic aspects of de Sitter entanglement entropy.

The static coordinatisation of de Sitter is interesting in the sense that the cosmological event horizon is explicitly `visible' in it and second,
it is explicitly time translational
invariant (within the cosmological event horizon), e.g.~\cite{Gibbons:1977mu}. The maximal analytic extension of the spacetime across this horizon shows, alike that of the non-extremal black hole or the Rindler horizon, four causally disconnected spacetime regions, two of which are static, \ref{review}.  A quantum field living in these two regions possesses local vacua  as well as some global vacuum defined with respect to the Kruskal null coordinates. It is thus interesting to address the issue of quantum entanglement between the quantum fields living in these two static regions, say R and L.  This issue was addressed recently in~\cite{Higuchi:2018tuk} for a scalar field, using the closed form mode functions and the behaviour of the entanglement entropy was shown to be similar to that of the Rindler spacetime. 

In this work we wish to do the same for the Dirac fermions. Precisely, using the closed form mode functions obtained in \ref{Dirac} for the static de Sitter coordinate,~\ref{st1}, we derive the R-L entanglement for the vacuum state of the Dirac field in \ref{global}. The result is found to be similar to that of the Rindler spacetime,  found earlier in~\cite{Alsing:2006cj, Mann:2009}. Such similarity follows from the universal Rindler-like behaviour of the $t-r$-part of any (no-extremal) near  Killing horizon metric  and the subsequent simplification of the mode functions. Such universality and simplification allow us to extend our result further to a) the de Sitter horizon of a general static and spherically symmetric spacetime, such as the Schwarzschild-de Sitter, \ref{global} and also to b) stationary axisymmetric spacetime such as the Kerr-de Sitter, \ref{kds}. For the latter in particular, we numerically investigate the behaviour of the entanglement entropy with respect to the variation of the energy, the angular momentum eigenvalues as well as the spacetime mass and rotation parameters.

We shall work with the mostly negative signature of the metric in $3+1$-dimensions and will set $c=G=\hbar=1$ throughout.

	\section{The static de Sitter -- a quick review}\label{review}
	The metric of the static patch of the de Sitter spacetime reads,
\begin{eqnarray}
ds^2 = (1-r^2) dt^2 -(1-r^2)^{-1} dr^2 - r^2 \left(d\theta^2 + \sin^2 \theta\, d\phi^2 \right)
\label{st1}
\end{eqnarray}
where the radial coordinate is made dimensionless via scaling by $H_0^{-1}=\sqrt{3/\Lambda}$. The coordinate system is not well defined for $r\geq 1$. This corresponds to the fact that the timelike Killing vector field of \ref{st1} only exists in $0\leq r < 1$. The $r=1$ null hypersurface is the cosmological event horizon serving as the causal boundary of our universe. 
	
The singularity at $r=1$ of this metric can be removed by choosing Kruskal-like coordinates which analytically continues the metric into the region $r\geq 1$, e.g.~\cite{Gibbons:1977mu}. In terms of the Kruskal null coordinates ($\overline{u}, \overline{v}$), the metric reads
\begin{eqnarray}
ds^2 = (1+r)^2 d \overline{u} d  \overline{v} - r^2 \left(d\theta^2 + \sin^2 \theta\, d\phi^2 \right)
\label{st2}
\end{eqnarray}
where $r$ above as a function of $\overline{u}, \overline{v}$ is understood and 
\begin{eqnarray}
&&\overline{u}: = - e^{u},  \qquad \overline{v}: =  e^{-v}\nonumber\\
&&u = t -r_{\star},   \qquad   v= t+r_{\star}, \qquad 
r_{\star} = \frac12 \ln \Big \vert \frac{1+r}{1-r} \Big \vert 
\label{st3}
\end{eqnarray}
The tortoise coordinate $r_{\star}$ reaches $\infty$ as $r\to 1$. The metric \ref{st2} has no singularity at $r=1$.

\begin{figure}[h!]
  \centering
    \includegraphics[width=4 cm, height = 4 cm]{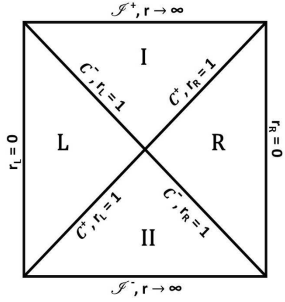}
 \caption{ The Penrose diagram of the de Sitter spacetime, found after the maximum analytic extension of \ref{st1}. Each point of the above diagram is a 2-sphere.  ${\cal C}^{\pm}$ denote respectively the future and past cosmological event horizons.  The causally disconnected static patches R and L are the regions of our concern. Region L is the time reversal of R and hence the timelike Killing vector field is taken to be past directed in L.  Regions I and II are non-static located in $r>1$. The infinities,  ${\mathscr{I}}^{\pm}$, are spacelike.    }
		\label{fig1}
\end{figure}

The coordinate system described in \ref{st2} and \ref{st3} covers the region R $(0\leq r_R \leq 1)$  of \ref{fig1}. ${\cal C^{\pm}}$ are respectively the future and past segments of the cosmological event horizon. We have 
\be
 \overline {u} ({\cal C}^+ )=0 = \overline {v} ({\cal C}^- ) 
 \label{C}
 \ee
and in general,
$$\overline {u} \leq 0, \quad \overline {v} \geq 0 \qquad {\rm (Region ~R)}$$
Likewise in region L, we define	
\begin{eqnarray}
\overline{u}: =  e^{u},  \qquad \overline{v}: = - e^{-v}
\label{st4}
\end{eqnarray}
so that  
 $$\overline {u} \geq 0, \quad \overline {v} \leq 0 \qquad {\rm (Region ~L)}$$
with \ref{C} still holding.

Regions I and II, being located at $r>1$, are not endowed with any timelike Killing vector field. We have 
$$\overline {u} > 0, \quad \overline {v} > 0 \quad {\rm (Region ~I)} \qquad \overline {u} < 0 ,\quad \overline {v} <0 \quad {\rm (Region ~II)}$$
It is manifest from the orientations of ${\cal C}^{\pm}$ in \ref{fig1} that the directions of the timelike Killing vector field, $\partial_t$, are opposite with respect to each other in R and L. We shall accordingly take it to be future directed in R, whereas past directed in L. The existence of ${\cal C}^{\pm}$, just like the black hole, splits the spacetime into four causally disconnected wedges R, L, I and II. For example, a particle initially in  R can only cross ${\cal C}^+$ to enter region I along a future directed trajectory, whereas it can only enter R from the contracting region L via ${\cal C}^-$.

 Similar causally disconnected regions also show up across the cosmological event horizon of a general static spherically symmetric spacetime (e.g. the Schwarzschild-de Sitter) and the stationary axisymmetric spacetime (e.g. the Kerr-de Sitter). For all of them, we shall investigate the quantum entanglement between regions R and L for a free massive Dirac field. 
	
\section{The Dirac equation in the static  de Sitter spacetime} \label{Dirac}

We shall solve in this section for the four mode functions of a free massive Dirac field in a closed form, in the background of \ref{st1} for both regions R and L in \ref{fig1}. The Latin indices appearing below will stand for the local Lorentz frame whereas the Greek indices will denote  the curved spacetime.

The Dirac equation in the static de Sitter coordinates was previously studied in~\cite{Otchik:1985ih, LopezOrtega:2006ig, LopezOrtega:2007sr} (also references therein) using the Newman-Penrose formalism for spinors developed in~\cite{Chandrasekhar}. See also~\cite{Cotaescu:1998ay, Cotaescu:2007xv} for discussions on choices of spacetime bases and group theoretic treatment of Dirac operators in static and time dependent de Sitter coordinates.

The Dirac equation in a  generally covariant form reads, 
	\begin{eqnarray}
	\left[i \gamma^\mu D_\mu-m_0\right]\Psi=0.
	\label{DE}
	\end{eqnarray}
where $m_0$ is the rest mass of the field. $\gamma_{\mu}:= e_{\mu}{}^a\gamma_a$, where $e_{\mu}{}^a$'s are the tetrad. The spin covariant derivative is defined as
	\begin{eqnarray}
	D_\mu:=\partial_\mu+\frac{1}{2}\omega_{\mu a b}\Sigma^{ab}
	\end{eqnarray}
where $\Sigma^{ab}=\left[\gamma^a,~\gamma^b\right]/4$ and the Ricci rotation coefficients $\omega$'s are given by : 
 $$\omega_\mu{}^a{}_b = - e_b{}^\nu \left(\partial_\mu e^a{}_\nu - \Gamma^\lambda_{\mu \nu} e^a{}_\lambda \right)$$
Following~\cite{wheeler}, we now define for \ref{st1}
\begin{eqnarray*}
	e^{~~ \mu}_a \equiv \  \left(
	\begin{array}{cccc}
		\left(1 - r^2\right)^{-\frac{1}{2}} & 0 & 0 & 0 \\ \vspace{2 mm}
		0 & \left(1 - r^2\right)^{\frac{1}{2}} \sin \theta  \cos \phi & \left(1 - r^2\right)^{\frac{1}{2}} \sin \theta  \sin \phi & \left(1 - r^2\right)^{\frac{1}{2}} \cos \theta  \\  \vspace{2 mm}
		0 & r^{-1} \cos \theta \cos \phi & r^{-1}   \cos \theta \sin \phi & - \ r^{-1}   \sin \theta \\  \vspace{2 mm}
		0 & - \left(r \sin \theta \right)^{-1}  \sin \phi &   \left(r \sin \theta \right)^{-1} \cos \phi & 0  \\
	\end{array}\right)
\end{eqnarray*}
\\
The Dirac equation is expanded to be,
\begin{eqnarray}
	\left[ \left(1 - r^2\right)^{-\frac{1}{2}} \left(\gamma^0 \ \partial_t - \dfrac{r}{2} \  \vec{\gamma} . \hat{r} \right) + \sqrt{1 - r^2} \ \left( \vec{\gamma} \cdot \hat{r} \right) \ \partial_r     - \left( \dfrac{1 - \sqrt{1 - r^2} }{r} \right) \ \vec{\gamma} . \hat{r} + \dfrac{1}{r} \left( \vec{\gamma} . \hat{\theta} \ \partial_\theta + \dfrac{\vec{\gamma} . \hat{\phi}}{\sin \theta} \ \partial_\phi   \right)          \right]    \Psi  \ + i m_0 \Psi=0,
	 \nonumber\\ 
	\label{DE2}
\end{eqnarray}
where $\hat{r}, \hat{\theta}, \hat{\phi}$ refer to the usual unit vectors in spherical polar coordinates and the gamma matrices are defined as,
\begin{eqnarray*}
\gamma^0=\left(
\begin{array}{cc}
I & 0\\
0 & -I
\end{array}
\right),\qquad
\gamma^i=\left(
\begin{array}{cc}
0 & \sigma^i\\
-\sigma^i & 0
\end{array}
\right), \,\,\,\, i=1,2,3.
\end{eqnarray*}
Decomposing $\Psi$ as
 $$\Psi = \begin{pmatrix} \Psi_1 \\ \Psi_2  \end{pmatrix} $$
 where $\Psi_1$ and $\Psi_2$ are each two component spinors, and using 
  $$\left( \vec{\sigma} . \hat{\theta} \ \partial_\theta + \dfrac{\vec{\sigma} . \hat{\phi}}{\sin \theta} \ \partial_\phi   \right) = - \left(  \vec{\sigma} . \hat{r} \right) \left(\hat{K_2} - 1 \right) $$
   where $\hat{K_2} =  \vec{\sigma} . \hat{L}  + 1 $, is an eigenoperator of the spherical spinors, we find that \ref{DE2} splits into two coupled equations, 
\begin{eqnarray}\label{coupled1}
\left( \left(1 - r^2\right)^{-\frac{1}{2}}  \partial_t + i m_0 \right) \Psi_1 + \left( \vec{\sigma} \cdot \hat{r} \right) \left(- \dfrac{r \left(1 - r^2\right)^{-\frac{1}{2}}}{2} + \sqrt{1 - r^2}\,\, \partial_r + \dfrac{\sqrt{1 - r^2}}{r} - \dfrac{\hat{K_2}}{r} \right)  \Psi_2  \ = 0 \nonumber \\
\left( \left(1 - r^2\right)^{-\frac{1}{2}}  \partial_t - i m_0 \right) \Psi_2 + \left( \vec{\sigma} \cdot \hat{r} \right) \left(- \dfrac{r \left(1 - r^2\right)^{-\frac{1}{2}}}{2} + \sqrt{1 - r^2}\,\, \partial_r + \dfrac{\sqrt{1 - r^2}}{r} - \dfrac{\hat{K_2}}{r} \right)  \Psi_1  \ = 0
\end{eqnarray} 

We recall that the spherical spin-1/2 harmonics, $\Omega_{jlm}$, are simultaneous eigenfunctions of $L^2, S^2, J^2, J_z$, given by~\cite{lifshitz},
\begin{eqnarray*}
	\Omega_{l+1/2, l , m}(\theta, \phi)=   \left(
	\begin{array}{c}
		C_{lm}^{+}\,  Y_{l, m - 1/2}(\theta, \phi) \\
		C_{lm}^{-}\,  Y_{l, m + 1/2}(\theta, \phi) 
	\end{array}
	\right), \qquad
	\Omega_{l-1/2,l,m}(\theta, \phi)=   \left(
	\begin{array}{c}
		- C_{lm}^{-}\,Y_{l, m - 1/2}(\theta, \phi) \\
	C_{lm}^{+}\, Y_{l, m + 1/2} (\theta, \phi)
	\end{array}
	\right)
\end{eqnarray*}
where
$$C_{lm}^{\pm}=\sqrt{\frac{l \pm m + 1/2}{2 l + 1}}$$

They satisfy the orthonormality relations (with $j=l\pm 1/2$)
\begin{eqnarray}\label{spherical-orthonormality}
\int \sin\theta\, d\theta\, d\phi \ \Omega_{j,l,m}^{\dagger}\, \Omega_{j',l',m'} = \delta_{jj'} \delta_{ll'} \delta_{mm'}
\end{eqnarray}
Further, we shall make use of the following easily verifiable properties,
\begin{eqnarray}
&&\left( \vec{\sigma} \cdot \hat{r} \right) \Omega_{l+1/2, l , m}   = -   \Omega_{(l+1)-1/2, l+1 , m}, \qquad \hat{K_2}  \Omega_{l+1/2, l , m}   =  (l+1)  \Omega_{l+1/2, l , m} \nonumber \\
&&\left(\vec{\sigma} \cdot \hat{r} \right)  \Omega_{(l+1)-1/2, l+1 , m}   =  -  \Omega_{l+1/2, l , m}, \qquad\,\,\,
\hat{K_2} \Omega_{(l+1)-1/2, l+1 , m}   =  - (l+1)  \Omega_{(l+1)-1/2, l+1 , m}
\label{identities}
\end{eqnarray}		
%
\subsection{The mode functions in $R$ and $L$}	\label{local}
We now make the following ansatz of variable separation,
\begin{eqnarray}\label{mode1}
\Psi = \begin{pmatrix} \Psi_1 \\ \Psi_2  \end{pmatrix} = \ e^{- i p t} \ \dfrac{\left(1 - r^2\right)^{-\frac{1}{4}}}{r}  \begin{pmatrix} i g(r) \ \Omega_{l+1/2, l , m} \\ f(r) \ \Omega_{(l+1)-1/2, l+1 , m}  \end{pmatrix}
\end{eqnarray}
with $p>0$. Note that the future directedness of the timelike Killing vector field in R guarantees the above mode to be positive frequency. However, since that vector field is past directed in L, the above mode will be negative frequency in L. 

We insert \ref{mode1} into \ref{coupled1} and use \ref{identities} to obtain,
\begin{eqnarray}\label{coupled2}
\left(1 - r^2 \right)  f'  + \dfrac{(l+1)}{r} \sqrt{1 - r^2} f &=& \left(p - m_0 \sqrt{1 - r^2}  \right) g \nonumber \\
\left(1 - r^2 \right)  g'  - \dfrac{(l+1)}{r} \sqrt{1 - r^2} g &=& - \left(p + m_0 \sqrt{1 - r^2}  \right) f
\end{eqnarray}	
where a `prime' denotes differentiation once with respect to $r$. Following~\cite{LopezOrtega:2006ig}, we now write 
\begin{eqnarray}\label{e1}
f = \dfrac{R_1 - R_2}{2}, \ \ g = \,\dfrac{i(R_1 + R_2)}{2 } 
\end{eqnarray}	
Further defining,
\begin{eqnarray}\label{e2}
R_1 (r) = \left(1 - r^2 \right)^{- \frac{1}{4}} \sqrt{1 - r} \ \left(f_1(r) - f_2(r) \right), \quad 
R_2 (r) = \left(1 - r^2 \right)^{- \frac{1}{4}} \sqrt{1 + r} \ \left(f_1(r) + f_2(r) \right) 
\end{eqnarray}	
\ref{coupled2} can be expressed as
\begin{eqnarray}\label{coupled 3}
\left(1 - r^2 \right) f'_1  - \left( \dfrac{l+1}{r} -i m_0 r\right)f_1  = - \left[\dfrac{1}{2} + i p - (l+1) + i m_0  \right] f_2       \nonumber \\
\left(1 - r^2 \right) f'_2  + \left( \dfrac{l+1}{r} -i m_0 r\right) f_2 = - \left[\dfrac{1}{2} + i p + (l+1) - i m_0  \right] f_1
\end{eqnarray}
 Form the above equation we get an uncoupled equation for $f_1$,
\begin{eqnarray}
\left(1-r^2\right)^2 f_1'' -2 r \left(1-r^2\right) f_1'+ \left[\left(1-r^2\right) \left(\frac{l+1}{r^2}+i m_0\right)   -\left(\frac{l+1}{r}-i m_0 r\right)^2+(l-i m_0+1)^2-\left(\frac{1}{2}+i p\right)^2\right] f_1 =0 \nonumber\\
\end{eqnarray}	
The solution that is regular as $r\to 1$ is given by, 
\begin{eqnarray}\label{e3}
f_1(r) = r^{l+1} \left(1-r^2\right)^{\frac{1}{4}+\frac{i p}{2}} \, F\left(\frac{ (2 l+2 i m_0+2 i p+5)}{4},\frac{(2 l-2 i m_0+2 i p+3)}{4};\frac{3}{2}+i p;1-r^2\right)
\end{eqnarray}	

We next plug $f_1$ into \ref{coupled 3} to determine $f_2$. Using the identities 15.2.15, 15.2.24 and 15.2.17 of~\cite{abramowitzstegun}, we find
$$  F(a,b+1;c;z) - \dfrac{a z}{c} \ F(a+1,b+1;c+1;z)  - F(a,b;c;z) = 0 $$
which yields,
\begin{eqnarray}\label{e4}
f_2(r) = - \dfrac{(2 l-2 i m_0+2 i p+3)}{(2 l-2 i m_0-2 i p+1)} \, r^{l+2} \left(1-r^2\right)^{\frac{1}{4}+\frac{i p}{2}}  \,F\left(\frac{(2 l+2 i m_0+2 i p+5)}{4},\frac{(2 l-2 i m_0+2 i p+7)}{4};\frac{3}{2}+i p;1-r^2\right)  \nonumber \\
\end{eqnarray}
Denoting now the mode function~\ref{mode1} by $\Psi^{R1+}_{plm}$, in terms of $f_1(r)$ and $f_2(r)$ we have
\begin{eqnarray}
\Psi^{R1+}_{p l m}  = \ e^{- i p t} \ \dfrac{\left(1 - r^2\right)^{-\frac{1}{2}}}{2~ r}  \begin{pmatrix}  -\left[\sqrt{1 - r} \ \left(f_1(r) - f_2(r) \right) + \sqrt{1 + r} \ \left(f_1(r) + f_2(r) \right)  \right] \Omega_{l+1/2, l , m} \\ \left[ \sqrt{1 - r} \ \left(f_1(r) - f_2(r) \right) - \sqrt{1 + r} \ \left(f_1(r) + f_2(r) \right)  \right] \Omega_{(l+1)-1/2, l+1 , m} 
 \end{pmatrix}
 \label{modep1}
\end{eqnarray}

A few comments on the normalisability of the mode functions are in order here. In~\cite{Otchik:1985ih, LopezOrtega:2006ig, LopezOrtega:2007sr}, the Dirac mode found is bounded at the origin, $r\to 0$, whereas they are seemingly not well defined at the cosmological horizon, $r=1$. Accordingly, the normalisation integral for those modes must exclude the horizon. This problem is present for a massive scalar and a vector field as well, as pointed out in~\cite{Higuchi:1986ww}. Our mode function, \ref{modep1}, on the other hand is not well behaved as $r\to 0$, follows from the properties of the hypergeomeric function~\cite{abramowitzstegun}. Accordingly, its normalisation integral must exclude $r=0$, which can be realised by considering modes which are localised near the horizon.  To the best of our knowledge, a Dirac mode in the static de Sitter coordinate  that is regular everywhere in $0\leq r \leq 1$ is unknown.  Such problem may just be a coordinate artefact.   The mode function we have taken will be appropriate to make the analytic continuation across the horizon in order to form the global modes, for our current purpose.
  \\

In order to find out the second positive frequency mode, we make the ansatz by `flipping' the angular part of the previous ansatz \ref{mode1},
\begin{eqnarray}\label{mode2}
\Psi^{R2+}_{pl m}  = \ e^{- i p t} \ \dfrac{\left(1 - r^2\right)^{-\frac{1}{4}}}{r}  \begin{pmatrix} i \zeta_1(r)   \Omega_{(l+1)-1/2, l+1 , m}  \\ \zeta_2(r)  \hspace{7 mm} \Omega_{l+1/2, l , m}  \end{pmatrix}
\end{eqnarray}
for which	\ref{coupled1} takes the form,
\begin{eqnarray} 
\left(1 - r^2 \right) \zeta_1'  + \dfrac{(l+1)}{r} \sqrt{1 - r^2} \ \zeta_1 &=& - \left(p + m_0 \sqrt{1 - r^2}  \right) \zeta_2\nonumber \\
\left(1 - r^2 \right)  \zeta_2'  - \dfrac{(l+1)}{r} \sqrt{1 - r^2} \ \zeta_2 &=& \left(p - m_0 \sqrt{1 - r^2}  \right) \zeta_1
\label{positive2}
\end{eqnarray}
Comparing the above with \ref{coupled2}, we find that the equations become identical if we let $f\to -\zeta_1, \, g \to \zeta_2$ and $m_0 \to -m_0$. This  determines  $\zeta_1(r)$ and $\zeta_2(r)$ in terms of $f_1$ and $f_2$ in \ref{e3}, \ref{e4}, with $m_0$ replaced with $-m_0$.\\

 \bigskip
 
Having thus determined the two positive frequency modes, we can obtain the negative frequency ones, simply found via the charge conjugation.  Given a positive frequency mode $u_p$, its charge conjugation $v_p$ is defined as,
\begin{eqnarray}\label{charge-conjugation}
v_p = C \left(\overline{u}_p\right)^T = i \gamma^2 u^{\star}_p. 
\end{eqnarray}
Using the explicit expressions of the spherical spinors appearing below \ref{coupled1}, it is easy to check that
\begin{eqnarray*}
	 \left( i \sigma_2 \right)  \Omega^*_{l+1/2, l , m}   = (-1)^{m + \frac{1}{2}}  \Omega_{l+1/2, l , -m}, \quad {\rm and} \quad
	\left( i \sigma_2 \right)  \Omega^*_{l-1/2, l , m}   =  (-1)^{m + \frac{3}{2}}  \Omega_{l-1/2, l , -m}. 
\end{eqnarray*}
Using this, from \ref{modep1} and \ref{mode2}, we can at once find out the two negative frequency modes, $\Psi^{(1-)}_{\omega l m}$, $\Psi^{(2-)}_{\omega l m}$. \\

For our purpose, we shall now demonstrate the  near cosmological horizon limit of these mode functions. We first recall that,
$$_2F_1\left(a,b,c,0\right)=1$$
Thus \ref{e3}, \ref{e4}  behave near the horizon as,
$$
f_1(r\to 1) \approx   (1-r^2)^{\frac14 + \frac{ip}{2}},  \qquad f_2(r\to 1) \approx  -\dfrac{(2 l-2 i m_0+2 i p+3)}{(2 l-2 i m_0-2 i p+1)} (1-r^2)^{\frac14 + \frac{ip}{2}},
$$
Also in terms of the tortoise coordinates \ref{st3}, we have as $r\to 1$, 
 $$\left(1-r^2\right)^{\frac{i p}{2}} = 2^{\frac{ip}{2}}\,e^{-ip r_{\star}}$$
Using these, and after normalising, we find from \ref{modep1}, \ref{mode2}, \ref{positive2} and \ref{charge-conjugation}
the expected plane wave solutions near the horizon in the R region,   
\begin{eqnarray}
&&\Psi^{R1+}_{{plm}} =  \ \frac{e^{- i p v_R} }{{\sqrt {4\pi} r} \left(1 - r^2\right)^{\frac14}  }\begin{pmatrix}   \Omega_{l+1/2, l , m} \\ \Omega_{(l+1)-1/2, l+1 , m}  \end{pmatrix}, \quad \Psi^{R2+}_{{plm}} =  \ \frac{ e^{- i p v_R} }{{\sqrt {4\pi} r} \left(1 - r^2\right)^{\frac14}  }\begin{pmatrix}   \Omega_{(l+1)-1/2, l+1 , m}  \\  \Omega_{l+1/2, l , m}  \end{pmatrix} \nonumber\\
&&\Psi^{R1-}_{{plm}} =   \frac{e^{ i p v_R} }{ {\sqrt {4\pi} r}\left(1 - r^2\right)^{\frac14}  }\begin{pmatrix} \Omega_{(l+1)-1/2, l+1 , -m}  \\  \Omega_{l+1/2, l , -m}  \end{pmatrix}, \quad \Psi^{R2-}_{{plm}} =  \frac{e^{ i p v_R} }{{\sqrt {4\pi} r} \left(1 - r^2\right)^{\frac14}  }\begin{pmatrix}  \Omega_{l+1/2, l , -m} \\  \Omega_{(l+1)-1/2, l+1, -m} 
\end{pmatrix} 
\label{NHR}
\end{eqnarray}
where in order to distinguish the region we have explicitly put the level `R'  and $v_R= t_R+r_{\star\, R }$. The  massless plane wave behaviour of the modes in the near horizon limit is consistent with the universal property of spacetimes endowed with a Killing horizon,~e.g.~\cite{Chandrasekhar}.  The overall factor $\sqrt{4\pi}$ in the denominators appearing above is necessary to ensure the normalisation, as described below.   \\

The inner product between any two modes $\Psi_i$ and $\Psi_j$ in the background of \ref{st1} is defined as,
$$(\Psi_i, \Psi_j)= \int \frac{r^2\, dr}{(1-r^2)^{\frac12}}\, \sin \theta\, d\theta\, d\phi \,\Psi_i^{\dagger}\,\Psi_j$$
Since the mode functions we have found are assumed to be localised near ${\cal C^{\pm}}$, and the inner product is independent of time, we may take the constant time hypersurface of the above integration to be infinitesimally close  to either ${\cal C^{\pm}}$ in \ref{fig1}, where the modes take the simple form, \ref{NHR}. We choose our surface to be close to ${\cal C^{+}}$. We have, for example 
\begin{eqnarray*}
(\Psi^{R1+}_{{plm}}, \Psi^{R1+}_{{p'l'm'}})=&&
\int_{r\to 1} \frac{r^2\, dr}{(1-r^2)^{\frac12}}\, \sin \theta\, d\theta\, d\phi \,\left(\Psi^{R1+}_{{plm}}\right)^{\dagger} \Psi^{R1+}_{{p'l'm'}} \nonumber\\= &&\delta{jj'}\delta_{ll'}\delta_{m m'} \int_{-\infty}^{\infty} \frac{dv_R}{2\pi}\, e^{-i (p-p') v_R} =  \delta(p-p')\, \delta{jj'}\,\delta_{ll'}\,\delta_{m m'}
\end{eqnarray*}
where we have used \ref{st3} and also \ref{spherical-orthonormality}. 

Likewise we can prove that the rest of the mode functions appearing in \ref{NHR} are normalisable and the inner product between any two different modes is vanishing. Thus \ref{NHR}  form a complete orthonormal set  in  R.   \\

The orthonormal modes in L  can be found in exactly the same manner as described above. However,  since the timelike Killing vector field is past directed there, the function $e^{ip t}$ should behave as positive frequency.  We shall write down the near horizon forms only,
\begin{eqnarray}
&&\Psi^{L1+}_{{plm}} =  \ \frac{e^{i p v_L} }{{\sqrt {4\pi} r} \left(1 - r^2\right)^{\frac14}  }\begin{pmatrix}   \Omega_{l+1/2, l , m} \\ \Omega_{(l+1)-1/2, l+1 , m}  \end{pmatrix}, \quad \Psi^{L2+}_{{plm}} =  \ \frac{e^{i p v_L} }{{\sqrt {4\pi} r} \left(1 - r^2\right)^{\frac14}  }\begin{pmatrix}   \Omega_{(l+1)-1/2, l+1 , m}  \\  \Omega_{l+1/2, l , m}  \end{pmatrix} \nonumber\\
&&\Psi^{L1-}_{{plm}} =   \frac{e^{ -i p v_L} }{{\sqrt {4\pi} r} \left(1 - r^2\right)^{\frac14}  }\begin{pmatrix} \Omega_{(l+1)-1/2, l+1 , -m}  \\  \Omega_{l+1/2, l , -m}  \end{pmatrix}, \quad \Psi^{L2-}_{{plm}} =  \frac{e^{ -i p v_L} }{{\sqrt {4\pi} r} \left(1 - r^2\right)^{\frac14}  }\begin{pmatrix}  \Omega_{l+1/2, l , -m} \\  \Omega_{(l+1)-1/2, l+1 , -m} 
\end{pmatrix} 
\label{NHL}
\end{eqnarray}

We can likewise find out the complete orthonormal set of R-L modes characterised by the retarded coordinate $u$.

In the following, we shall use the causally disconnected near horizon local mode functions, \ref{NHR}, \ref{NHL},  to construct the global modes and will compute the Bogoliubov coefficients.

\section{The global modes and the Bogoliubov coefficients}\label{global}

We can construct the global modes having support in $R\cup L$ in a manner similar to that of the Rindler spacetime, e.g.~\cite{Unruh:1976db, Socolovsky:2013rga}. We shall analytically continue an R mode to L and vice versa along a complex path, by using the Kruskal coordinates, well defined on ${\cal C^{\pm}}$ as discussed in \ref{review}. Let us take the pair $(\Psi^{R1+}_{{plm}}, \Psi^{L2-}_{{plm}})$ from \ref{NHR}, \ref{NHL}. Since both of these modes behave as $e^{-ipv}$ near the horizon with $p>0$, we shall choose our complex path through the lower half plane.  The angular parts remain intact during this analytic continuation procedure. 

In terms of the Kruskal coordinates, \ref{review}, we have $\Psi^{R1+}_{{plm}}\sim (\overline{v})^{-ip}$ and $\Psi^{L2-}_{{plm}}\sim (-\overline{v})^{-ip}$. Thus while continuing the latter mode through the lower half plane onto the region R, we must write, $-\overline{v}= e^{-i\pi} \overline{v}$. Accordingly we have $\Psi^{L2-}_{{plm}}\sim e^{-\pi p} (\overline{v})^{-ip}$. Thus the linear combination,
$$ \Psi^{R1+}_{{plm}}+e^{-\pi p}\Psi^{L2-}_{{plm}}  $$
is analytic across the horizon, have support in $R\cup L$ and behaves as a global mode function. Normalising, we denote this mode function as
\begin{eqnarray}
_G\Psi^{(1)}_{plm} =  \frac{1}{\sqrt{2 \cosh p \pi}}  \left(e^{\frac{\pi p}{2}} \,\Psi^{R1+}_{{plm}} + e^{- \frac{\pi p}{2}} \Psi^{L2-}_{{plm}} \right) 
\label{g1}
\end{eqnarray}
Likewise, we find three other global mode functions to form an orthonormal set 
\begin{eqnarray}\label{globalmodes2}
&&_G\Psi^{(2)}_{plm}  = \dfrac{1}{\sqrt{2 \cosh p \pi}}  \left(e^{\frac{\pi p}{2}} \, \Psi^{R2+}_{{plm}} + e^{- \frac{\pi p}{2}}\Psi^{L1-}_{{plm}}\right) \nonumber \\
&&_G\Psi^{(3)}_{plm}  = \frac{1}{\sqrt{2 \cosh p \pi}}  \left(e^{\frac{\pi p}{2}} \,\Psi^{R1-}_{{plm}} - e^{- \frac{\pi p}{2}} \Psi^{L2+}_{{plm}} \right) \nonumber \\
&&_G\Psi^{(4)}_{plm}  = \frac{1}{\sqrt{2 \cosh p \pi}}  \left(e^{\frac{\pi p}{2}} \,\Psi^{R2-}_{{plm}} - e^{- \frac{\pi p}{2}} \Psi^{L1+}_{{plm}} \right)
\end{eqnarray}
Clearly, there exists another set of global modes, $_G\Psi^{(5)}_{plm},\, _G\Psi^{(6)}_{plm},\,_G\Psi^{(7)}_{plm},\,_G\Psi^{(8)}_{plm}$,  found via the interchange   $R \leftrightarrow L$ on the right hand side of \ref{g1} and \ref{globalmodes2}.\\

We quantise the Dirac field $\Psi$ now in $R\cup L$ in terms of the local modes, \ref{NHR}, \ref{NHL} as well as the global modes found above. In terms of the local modes, we have
\begin{eqnarray}
\label{localdirac}
\Psi=\sum_{l m s} \int_0^{\infty} dp
\left( c_{s_{ p l m}}^R \Psi^{Rs+}_{{plm}}+d_{s_{ p l m}}^{R\dag}  \Psi^{Rs-}_{{plm}} 
+c_{s_{ p l m}}^L\Psi^{Ls+}_{{plm}}+d_{s_{ p l m}}^{L\dag} \Psi^{Ls-}_{{plm}}  \right),
\end{eqnarray}
where $s= 1, 2$. The operators satisfy the usual anti-commutation relations
\begin{eqnarray}
\label{ac}
&&\left[c_{s_{p l m}}^R,~c_{{s'}_{ p' l' m'}}^{R \dag}\right]_+
=\left[d_{s_{p l m}}^R,~d_{{s'}_{ p' l' m'}}^{R \dag}\right]_+
=\delta\left(p-p'\right)\delta_{ss'}\delta_{ll'}\delta_{mm'} \nonumber\\
&&\left[c_{s_{p l m}}^L,~c_{{s'}_{ p' l' m'}}^{L \dag}\right]_+
=\left[d_{s_{p l m}}^L,~d_{{s'}_{ p' l' m'}}^{L \dag}\right]_+
=\delta\left(p-p'\right)\delta_{ss'}\delta_{ll'}\delta_{mm'}
\end{eqnarray}
and all the other anti-commutators vanish.  

Likewise in terms of  the global modes we have,
\begin{eqnarray}
\Psi=\sum_{l m}\int_0^{\infty} dp
\left(  a_{1 p l m}\, _G\Psi^{(1)}_{plm} +a_{2 p l m}\,  _G\Psi^{(2)}_{plm} +
 b_{ 1p l m}^{\dag}\, _G\Psi^{(3)}_{plm}+ b_{ 2p l m}^{\dag}\,  _G \Psi^{(4)}_{plm} \right.
  \nonumber\\  \left. +a_{3 p l m}\, _G\Psi^{(5)}_{plm} +a_{4 p l m}\,  _G\Psi^{(6)}_{plm} +
 b_{ 3p l m}^{\dag}\, _G\Psi^{(7)}_{plm}+ b_{ 4p l m}^{\dag}\,  _G \Psi^{(8)}_{plm} \right)
 \label{globalf}
\end{eqnarray}

Comparing  \ref{localdirac} and \ref{globalf} via \ref{g1}, \ref{globalmodes2}, we obtain the Bogoliubov relations,
\begin{eqnarray}\label{Bogoliubov}
a_{1plm} &=& \dfrac{1}{\sqrt{2 \cosh \pi p}} \left(e^{\frac{\pi p}{2}} \, c^L_{1_{plm}} - \, e^{- \frac{\pi p}{2}} \, d^{R\dag}_{2_{pl-m}} \right),
\quad a_{2plm} = \dfrac{1}{\sqrt{2 \cosh \pi p}} \left(e^{\frac{\pi p}{2}} \, c^L_{2_{plm}} + \, e^{- \frac{\pi p}{2}} \, d^{R\dag}_{1pl-m} \right) \nonumber \\
b^{\dag}_{1plm} &=& \dfrac{1}{\sqrt{2 \cosh \pi p}} \left(e^{\frac{\pi p}{2}} \, d^{L\dag}_{1_{pl-m}} - \, e^{- \frac{\pi p}{2}} \, c^{R}_{2_{plm}} \right), 
\quad b^{\dag}_{2plm} = \dfrac{1}{\sqrt{2 \cosh \pi p}} \left(e^{\frac{\pi p}{2}} \, d^{L\dag}_{2_{pl-m}} + \, e^{- \frac{\pi p}{2}} \, c^{R}_{1_{plm}} \right)
\end{eqnarray}
Another similar set is obtained by making the  $R \leftrightarrow L$ interchange on the right hand side of  the above equations, for the operators $a_3,\, a_4, \, b_3,\, b_4$ appearing in \ref{globalf}.  It is easy to check using \ref{ac} that the global operators satisfy the canonical anti-commutation relations.

Since the above Bogoliubov coefficients are spacetime independent, the Bogoliubov relations are valid away from the horizons as well, even though we used the near horizon forms of the modes to derive it. In other words, \ref{globalmodes2} will hold even if we are away from the horizon, with the local modes taking their non-trivial forms, e.g. \ref{modep1}.

Being equipped with these, we are now ready to compute the particle creation and the R-L entanglement for \ref{st1} for a free massive Dirac field. 

\subsection{The entanglement entropy}\label{entropy-ds} 
The local vacua $  |0\rangle_R,  |0\rangle_L$ are defined as, 
\begin{eqnarray}
\label{localvacuua}
c_{s_{p l m}}^R  |0\rangle_R = d_{s_{p l m}}^R   |0\rangle_R = 0,  \qquad c_{s_{p l m}}^L  |0\rangle_L = d_{s_{p l m}}^L  |0\rangle_L = 0 \qquad (s=1,2) 
\end{eqnarray}
whereas the global vacuum is defined as
\begin{eqnarray}
\label{globalvac}
a_{{\sigma p l m}} |0\rangle = b_{\sigma p l m} |0\rangle =0.
\end{eqnarray}
where $\sigma =1,2,3,4$. Hereafter we shall drop the indices $(p,l,m)$ on the operators for the sake of brevity.\\

From the Bogoliubov relations of the preceding section, it is clear that   the operators  $\left(a_1, a_2, b_3, b_4 \right)$ and $\left(a_3, a_4, b_1, b_2 \right)$ can be grouped into two sectors, anti-commuting trivially. The global vacuum can therefore be decomposed as $|0\rangle=|0\rangle^{(1)} \otimes |0\rangle^{(2)}$, where  $|0\rangle^{(1)}$ is annihilated by the first set of operators and  $|0\rangle^{(2)}$ by the second. Accordingly, we shall work with  $|0\rangle^{(1)}$ only, for $|0\rangle^{(2)}$ will yield identical results. 

The Bogoliubov relations imply a squeezed state relation of $|0\rangle^{(1)}$ with the local R-L vacua,
\begin{eqnarray}\label{squeezedstate}
|0\rangle^{(1)} = N
\exp\left(\sum_{s,s'=1, 2} \xi_{s s'}\,c^{L\dagger}_{s} d^{R\dagger}_{s'} \right)
|0\rangle_R^{(1)} \otimes |0\rangle_L^{(1)},
\label{vac}
\end{eqnarray}
where $\xi_{ss'}$'s are four complex numbers, $d_{1}^R  |0\rangle_R^{(1)} = d_{2}^R |0\rangle_R^{(1)} = 0$ and $c_{1}^L |0\rangle_L^{(1)} = c_{2}^L |0\rangle_L^{(1)} = 0$ and $N$ is the normalisation. Thus we may further write
\begin{eqnarray}
|0\rangle_R^{(1)} = |0_{d_1}\rangle_R \otimes  |0_{d_2}\rangle_R \equiv |0_{d_1}, 0_{d_2} \rangle_R, \qquad |0\rangle_L^{(1)} = |0_{c_1}\rangle_L \otimes  |0_{c_2}\rangle_L \equiv |0_{c_1}, 0_{c_2} \rangle_L
\end{eqnarray} 
where  $|0_{c_1}\rangle_L$, $|0_{c_2}\rangle_L$ are annihilated by  $c_1^L$ and $c_2^L$ respectively. Likewise $|0_{d_1}\rangle_R$, $|0_{d_2}\rangle_R$ are annihilated by $d_1^R$ and $d_2^R$ respectively.\\

Since $|0\rangle^{(1)}$ is annihilated by  $a_1, a_2, b_3, b_4$, it turns out that
\begin{eqnarray}
\xi_{11} = \xi_{22} = 0, \qquad \xi_{12} = - \xi_{21} = e^{- \pi p} 
\end{eqnarray}
We  now write \ref{squeezedstate} as
\begin{eqnarray}
|0\rangle^{(1)} = N \left[1 + e^{- \pi p} \left(- |0, 1 \rangle_R \,  |1, 0 \rangle_L + |1, 0 \rangle_R \,  |0, 1 \rangle_L  \right) - e^{- 2 \pi p} \, |1, 1 \rangle_R \,  |1, 1 \rangle_L  \right]
\label{gvac}
\end{eqnarray}
 where the normalisation $N$ is given by
 %
$$N = \dfrac{1}{\left(1 + e^{- 2 \pi p}\right)}$$
%

As a check of consistency, we may compute the expectation value of the local number operator in the global vacuum. We find from \ref{Bogoliubov}, for example
%
$${}^{(1)}\langle 0 | c_{1}^{L \dag} c_{1}^{L} | 0 \rangle^{(1)} = N^2   \left(e^{- 2 \pi p} + e^{- 4 \pi p}  \right) = \dfrac{1}{ e^{2  \pi p}+1}$$
%
showing fermionic `black body' distribution  with temperature $1/2\pi$.

The density operator corresponding to the global vacuum state \ref{gvac}, is given by $\rho = |0\rangle^{(1)}  {}^{(1)}\langle 0| $. Tracing out now the states of the region inaccessible to us (say L), we find the reduced density operator $\rho_R = {\rm Tr_L} \left(|0\rangle^{(1)}  {}^{(1)}\langle 0| \right)$. Its matrix representation is given by
\begin{eqnarray}\label{reduceddensity}
\rho_R \equiv N^2 \left(
\begin{array}{cccc}
1 & 0 & 0 & 0 \vspace{3mm}\\
0 & e^{- 2 \pi p} & 0 & 0 \vspace{3mm}\\
0 & 0 & e^{- 2 \pi p} & 0 \vspace{3mm}\\
0 & 0 & 0 & e^{- 4 \pi p}  \\
\end{array}\right)
\end{eqnarray}

Finally, we obtain the entanglement entropy for a single mode characterised by $p$,
\begin{eqnarray}
S(p) = -{\rm Tr}\left(\rho_R\ln\rho_R\right)=
2 \, \ln\left(1 + e^{-2 \pi p} \right) + \dfrac{4 \pi p}{1 + e^{2 \pi p}}
\label{eentropy}
\end{eqnarray}

Note that the above result is identical to that of the Rindler spacetime~\cite{Alsing:2006cj, Mann:2009}. This is not surprising, as the entanglement we are obtaining here is due to the existence of the R-L  regions created by the bifurcation surface ${\cal C^{\pm}}$ in \ref{fig1}. Now, it is well known that the $t-r$ part of any non-extremal near horizon geometry is similar to that of the Rindler. Thus we expect a universality of the Bogoliubov coefficients, \ref{Bogoliubov}, for {\it all} non-extremal Killing horizons, be it black hole or cosmological,
at least at the qualitative level. 

We shall further extend below the above result  for both static spherically symmetric and stationary axisymmetric spacetimes.

\subsection{General static spherically symmetric spacetimes}\label{sph}
We take the ansatz,
\begin{equation}
ds^2 = f(r) dt^2 - h(r) dr^2 - r^2 d \Omega^2
\label{st2'}
\end{equation}
 We assume that the above metric admits a cosmological event horizon, located (in a dimensionless unit as earlier) at $r=1$,
 %
$$f(r\to 1) \to 0,\quad h^{-1}(r\to 1) \to 0,$$
%
with $f h\,(r\to 1) $ is neither vanishing nor divergent. \ref{st2'} can represent, e.g. the Schwarzschild-de Sitter spacetime (i.e., $f(r)=(1-2M/r-H_0^2 r^2) = h^{-1}(r)$) or its charged variants.

The existence of the Killing horizon guarantees 
the existence of causally disconnected regions like R and L similar to \ref{fig1}. The tortoise coordinate is defined in this case as, 
\be
r_{\star}:= \int dr\,\sqrt{\dfrac{h}{f}},
\label{T}
\ee
whereas the Kruskal null coordinates at the cosmological event horizon will exactly be formally similar to
that of \ref{review}. Thus, as of \ref{local}, due to the spherical symmetry, we can look for a solution of the form,
%
$$\Psi = \begin{pmatrix} \Psi_1 \\ \Psi_2  \end{pmatrix} = \frac{e^{- i p t}} {r f^{\frac{1}{4}}}  \begin{pmatrix} i \tilde{\Psi}_1(r)  \Omega_{l+1/2, l , m} \\ \tilde{\Psi}_2(r)  \Omega_{(l+1)-1/2, l+1 , m}  \end{pmatrix}$$
%
so that we obtain
\begin{eqnarray}
\left( - \frac{p}{\sqrt f}  + m_0 \right) \tilde{\Psi}_1 +  \left( \frac{1}{\sqrt h} \partial_r + \dfrac{h^{-\frac{1}{2}}}{r} + \dfrac{(l+1)}{r} \right)  \tilde{\Psi}_2  = 0 \nonumber \\
\left( \frac{p}{\sqrt f}  + m_0 \right) \tilde{\Psi}_2 + \left( \frac{1}{\sqrt h} \partial_r + \dfrac{h^{-\frac{1}{2}}}{r} - \dfrac{(l+1)}{r} \right)  \tilde{\Psi}_1   = 0
\label{static-dirac}
\end{eqnarray} 

To the best of our knowledge, unlike the static de Sitter, the Dirac equation cannot be solved in a closed form, even in the Schwarzschild-de Sitter background. However we recall from the preceding section that in order to study the entanglement, we need the Bogoliubov relations between the local R-L modes and the global ones. In order to find the global modes on the other hand, we must analytically continue the local modes across  the horizon. Thus, we need to be concerned only about the near horizon forms of the modes.  

The near horizon limit of \ref{static-dirac} after using \ref{T} is given by,
\begin{eqnarray}
 \partial_{r_{\star}} \tilde{\Psi}_2   = p  \tilde{\Psi}_1,  \qquad
 \partial_{r_{\star}} \tilde{\Psi}_1   = - p \tilde{\Psi}_2
\end{eqnarray}
which yield modes exactly similar to \ref{NHR}, \ref{NHL}. By defining the Kruskal null coordinates analogously as the static de Sitter, we can find out the global modes similar to \ref{g1}, \ref{globalmodes2}. Hence we shall obtain identical Bogoliubov relations as \ref{Bogoliubov} and the entanglement entropy, \ref{eentropy}.

Note also that \ref{st2'} in addition can also possess a black hole event horizon, located inside the cosmological event horizon. As long as the black hole horizon is non-extremal, the computation of the  entanglement entropy for the black hole will be similar to that of the cosmological horizon. One needs two different sets of Kruskal coordinates to analyse the near horizons' modes.

We shall end this section by putting a comment of the Nariai limit of the Schwarzschild-de Sitter spacetime ($3\sqrt{3} M H_0\to 1$), for which the radial values of the cosmological and black hole  event horizons are nearly coincident. This makes the proper separation between them large.
Utilising this proper separation as a coordinate, the metric can be written as, e.g.~\cite{Nojiri:2013su},
\be
ds^2= \frac{1}{9H_0^4}\left[ \frac{1}{\cosh^2x} \left( dt^2 -dx^2 \right) - d \Omega^2\right]
\label{nariai}
\ee
The above metric is $dS^2\times S^2$ and is endowed with a black hole and cosmological event horizon, located respectively at $x \to \mp \infty$.  We can perform the above analysis in this case as well to obtain  similar results as \ref{eentropy}. 

\section{The stationary axisymmetric spacetimes}\label{kds}
We finally come to the case of the stationary axisymmetric spacetimes endowed with a positive $\Lambda$. For example, the Kerr-de Sitter  spacetime in the Boyer-Lindquist coordinates reads
\begin{eqnarray}
ds^2=\frac{\Delta_r-a^2\sin^2\theta \Delta_{\theta}}{\rho^2}dt^2 +\frac{2a \sin^2 \theta }{\rho^2 \Xi} \left( (r^2+a^2 )\Delta_{\theta}-\Delta_r\right)dt d\phi \nonumber\\ - \frac{\sin^2 \theta }{\rho^2 \Xi^2} \left( (r^2 +a^2)^2 \Delta_{\theta}-\Delta_r a^2 \sin^2\theta\right)d\phi^2 - \frac{\rho^2}{\Delta_r}dr^2 - \frac{\rho^2}{\Delta_{\theta}}d\theta^2
\label{sup1}
\end{eqnarray}
where,
\begin{equation}
\Delta_r = (r^2 +a^2) \left(1-H_0^2 r^2 \right) -2Mr, \quad \Delta_{\theta} =1 + H_0^2 a^2 \cos^2\theta,  \quad \Xi = 1+ H_0^2 a^2,  \quad \rho^2 =r^2 +a^2 \cos^2 \theta\,, 
\end{equation}
with $H_0^2 = \Lambda/3$ as earlier. The parameter  $a$ is related to the spacetime rotation.   For $a=0$,  we recover the Schwarzschild-de Sitter spacetime, whereas setting  $M=0$ we obtain the de Sitter spacetime written in the static patch.  

Unlike the previous cases, we shall retain the parameters intact, for the purpose of some numerical analysyes we wish to perform.   The cosmological event horizon is given by the largest root of $\Delta_r=0$. Its surface gravity is given by 
$$-\kappa_{C}= \frac{\Delta_r'}{2(r^2+a^2)}\bigg\vert_{r= r_C}$$
with $\kappa_C >0$.
Finally, the horizon Killing field is given by $\chi_{C}=\partial_t+\Omega_{C}\,\partial_{\phi}$, with 
$$\Omega_{C}= \frac{a\, \Xi}{r_{C}^2+a^2}$$
being the angular speed on the cosmological event horizon.  There can  be a black hole horizon at $r=r_H$ $(r_H \leq r_C)$, as well. 

Aspects of quantum entanglement in the Kerr spacetime for a scalar field with its various vacuum states can be seen in~\cite{Menezes:2017oeb}. Due to the existence of the ergosphere, the timelike Killing vector field in rotating spacetimes like \ref{sup1} becomes spacelike on or in the neighborhood of the horizon. In order to tackle this difficulty, usually one needs to consider the horizon Killing vector field, future directed and null on the horizon. This vector field describes a family of observers co-rotating with the same angular speed on the horizon.  Aspects of quantisation of fermions in  rotating backgrounds with the vacuum state defined with respect to such rigidly rotating observers can be seen in~\cite{Casals:2012es, Ambrus:2014uqa, Belokogne:2014ysa, Toth:2015cda}.

Now, the Dirac equation in \ref{sup1} can be studied using the usual Newman-Penrose basis, e.g.~\cite{Batic:2015wda} and references therein.  However, in order to define the aforementioned rigidly rotating states near the horizon, it is a bit convenient to go to a diagonal basis as follows. Let us define a vector field,
$$\chi^{\mu} = (\partial_t)^{\mu} -\frac{(\partial_t \cdot \partial_{\phi}) }{(\partial_{\phi}\cdot \partial_{\phi})}\,(\partial_{\phi})^{\mu}  = (\partial_t)^{\mu} -\frac{g_{t\phi} }{g_{\phi \phi}}\,(\partial_{\phi})^{\mu} \,$$
It satisfies $\chi\cdot \partial_\phi =0$ and also trivially $\chi\cdot \partial_r =0=\chi\cdot \partial_{\theta}$.  We have, 
$$\chi^{\mu}\chi_{\mu} =  \frac{g_{tt} g_{\phi\phi} - g^2_{t\phi}}{g_{\phi\phi}} = \frac{\rho^2 \Delta_r \Delta_\theta}{(r^2+a^2)^2 \Delta_\theta - \Delta_r a^2 \sin^2\theta}=\beta^2~({\rm say}) $$
which, with $\Delta_r>0$, is easily seen to be positive. In other words, $\chi^{\mu}$ is a timelike vector field and it is null on the horizon(s), $\Delta_r=0$. It can be also seen that a) $\chi^{\mu}$
satisfies the Frobenius condition of hypersurface orthogonality and b) even though $\chi^{\mu}$ is not in general Killing, it smoothly coincides with the horizon Killing field(s)~\cite{Bhattacharya:2011dq}. Thus we can define the following orthonormal basis for \ref{sup1},
$$ e_{0}^{\mu}=\beta^{-1}\chi^{\mu}, \quad e_{1}^{\mu}= \frac{1}{\sqrt{-g_{rr}}}(\partial_r)^{\mu},  \quad e_{2}^{\mu}= \frac{1}{\sqrt{-g_{\theta\theta}}}(\partial_{\theta})^{\mu}, \quad e_{3}^{\mu}= \frac{1}{\sqrt{-g_{\phi\phi}}}(\partial_{\phi})^{\mu}$$
With this choice and the representation of the $\gamma$-matrices mentioned in \ref{Dirac}, we expand \ref{DE}. For the positive frequency modes, we take the ansatz $\Psi= e^{-ipt+i m \phi}\, \widetilde{\Psi}(r,\theta) $. As earlier we focus only on the near horizon limit, $\Delta_r \to 0$,  to obtain
\begin{eqnarray}
&&\frac{(r^2+a^2)}{\rho \sqrt{\Delta_r}} \gamma^0 (p -m\Omega_C) \widetilde{\Psi} + \frac{i \sqrt{\Delta_r}}{\rho} \gamma^1 \partial_r \widetilde{\Psi} +\frac{i \sqrt{\Delta_{\theta}}}{\rho} \gamma^2 \partial_{\theta} \widetilde{\Psi} +\frac{i \rho\, \Xi}{(r^2+a^2)\sin \theta \sqrt{\Delta_{\theta}}} \gamma^3\partial_{\phi} \widetilde{\Psi}\nonumber\\
&&+\frac{i}{4} \left[ \frac{\Delta'_r}{\rho \sqrt{\Delta_r}}\gamma^1\gamma^2 + \frac{\rho \partial_{\theta} (\Delta_{\theta} \sin^2 \theta/\rho^2) }{\sin^2 \theta \sqrt{\Delta_{\theta}}} \right] \gamma^2 \,\widetilde{\Psi} -m \widetilde{\Psi}\, +\, {\cal O}(\Delta_r^2)=0
\label{kdsdirac}
\end{eqnarray} 
We multiply both sides with $\rho \sqrt{\Delta_r}/(r^2+a^2)$ and rewrite it using the tortoise coordinate defined as,
$$ dr_{\star} = \int \frac{(r^2+a^2)}{\Delta_r} $$
Note that as $r\to r_C$ or $\Delta_r \to 0$, various terms, including the mass term drop. We find for the two positive frequency near horizon modes in the region R,
\begin{eqnarray}
\Psi^{R1+}_{{p\lambda m}}=  \frac{\sqrt{\Xi}}{\sqrt{2\pi \rho}\,\Delta_r^{\frac14}}e^{- i (p-m\Omega_C) v_R +i m \phi_C}   \begin{pmatrix}  S_+(\lambda,\theta)\\ S_-(\lambda, \theta)\\ -S_-(\lambda, \theta)\\ S_+(\lambda, \theta)   \end{pmatrix},  \quad \Psi^{R2+}_{{p\lambda m}}=   \frac{\sqrt{\Xi}}{\sqrt{2\pi \rho}\,\Delta_r^{\frac14}}e^{- i (p-m\Omega_C) v_R +i m \phi_C }   \begin{pmatrix}  S_+(\lambda,\theta)\\ S_-(\lambda, \theta)\\ S_-(\lambda, \theta)\\ -S_+(\lambda, \theta)   \end{pmatrix} 
\label{kds-modes}
\end{eqnarray}
  where the untwisted azimuthal coordinate $\phi_C := \phi - \Omega_C t$, defines a rigidly rotating observer on the horizon and $v= t+r_{\star}$ as earlier.  The negative frequency modes are found via the charge conjugation of \ref{kds-modes}. The modes in the L region 
  are also found by making the time past directed, as in \ref{NHL}.  Note that in order to have any sensible field quantisation, we must have $p- m\Omega_C \geq 0$.
  
  The angular functions $S_{\pm}(\lambda, \theta)$ are spin-1/2 weighted spheroidal harmonics with eigenvalues $\lambda$, e.g.~\cite{Batic:2015wda} and references therein. They are normalised as,
 $$\int_{0}^{\pi} d\theta \sin \theta \left[S^{\star}_+(\lambda, \theta) S_+(\lambda', \theta)+ S^{\star}_-(\lambda, \theta) S_-(\lambda', \theta)\right] = \delta_{\lambda \lambda'}$$
We shall not require their explicit forms for our current purpose. 

Choosing our integration hypersurface to be orthogonal to the vector field $\chi^{\mu}$, it can be easily checked as earlier that the modes of \ref{kds-modes} (along with the negative frequency modes) are $\delta$-function normalisable and they form an orthonormal set.  With this, we now perform the same analysis as described in \ref{entropy-ds} to obtain a generalisation of \ref{eentropy},  
\begin{eqnarray}
S(p, m) = 2 \, \ln\left(1 + e^{-\frac{2 \pi (p- m \Omega_C)}{\kappa_C}} \right) + \dfrac{4 \pi (p-m\Omega_C)}{\kappa_C\left(1 + e^{\frac{2 \pi (p- m \Omega_C)}{\kappa_C}  } \right)}
\label{ee-kds}
\end{eqnarray}
Formally similar expression corresponding to the black hole event horizon is obtained simply by replacing $\kappa_C$ by $\kappa_H$ and $\Omega_C$ by $\Omega_H$, respectively the surface gravity and the angular speed of the black hole horizon. Setting  $\Omega_C=0$ in the above equation yields the result for the Schwarzschild-de Sitter spacetime. Setting further $M=0$ recovers \ref{eentropy}, with the dimensionless scaling $p \to p/\kappa_C$.\\
\begin{figure}
\centering
\begin{subfigure}{.45\textwidth}
  \centering
  \includegraphics[width=1\linewidth]{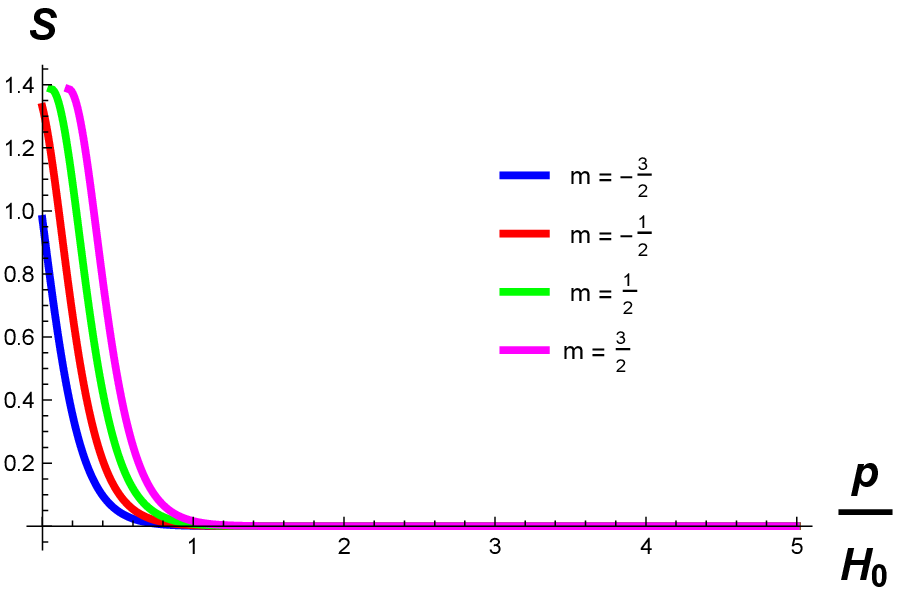}
  \caption{$a H_0 = 0.09$, $M H_0 = 0.1$}
\end{subfigure}%
\begin{subfigure}{.45\textwidth}
 \centering
  \includegraphics[width=1\linewidth]{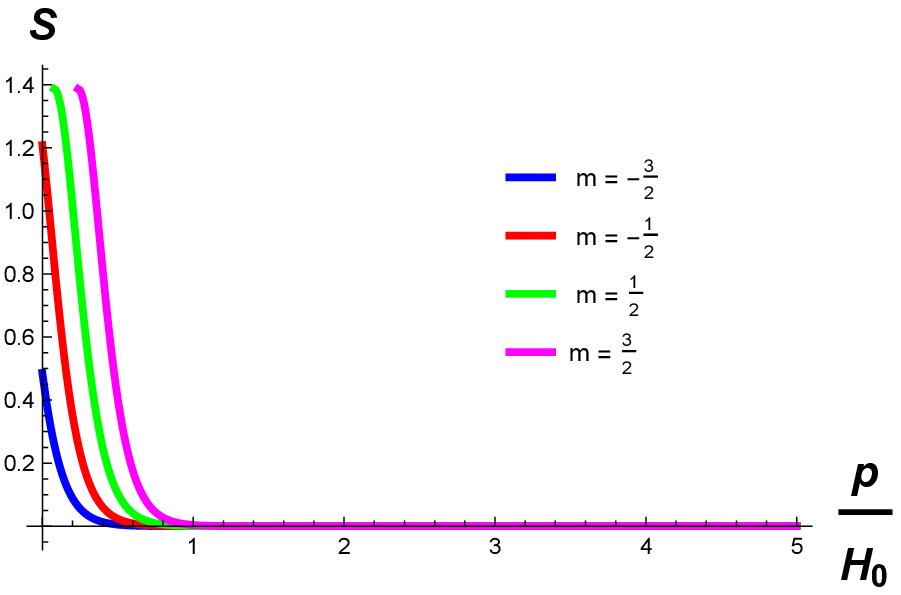}
  \caption{$a H_0 = 0.1$, $M H_0 = 0.15$}
\end{subfigure}
\caption{The variation of the entanglement entropy, \ref{ee-kds}, with respect to the dimensionless  energy $p/H_0$,  with different angular eigenvalues.  }
\label{fig2}
 \end{figure}
\begin{figure}
\centering
\begin{subfigure}{.45\textwidth}
  \centering
  \includegraphics[width=1\linewidth]{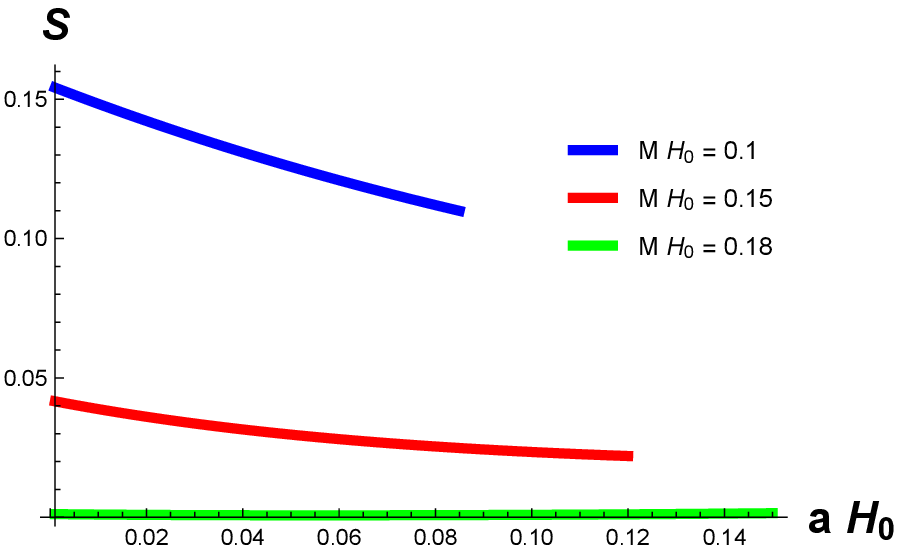}
  \caption{$ \frac{p}{H_0} = 0.5, m = -1/2$}
\end{subfigure}%
\begin{subfigure}{.45\textwidth}
 \centering
  \includegraphics[width=1\linewidth]{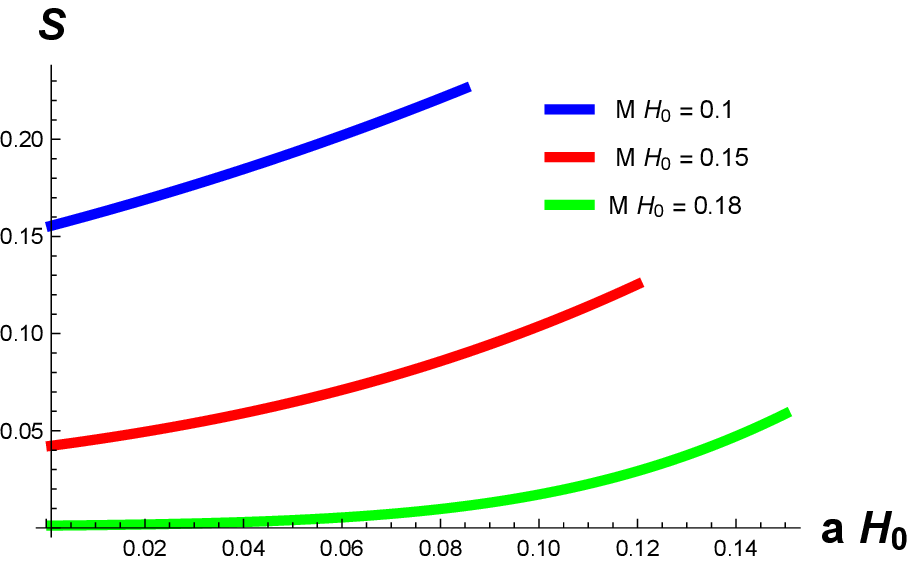}
  \caption{$ \frac{p}{H_0} = 0.5, m = 1/2$}
\end{subfigure}
\caption{The variation of the entanglement entropy, \ref{ee-kds}, with respect to the dimensionless rotation parameter, $aH_0$.  The qualitative difference of the $m>0$ and $m<0$ states is manifest here.}
\label{fig3}
\end{figure}
\begin{figure}
\centering
\begin{subfigure}{.45\textwidth}
  \centering
  \includegraphics[width=1\linewidth]{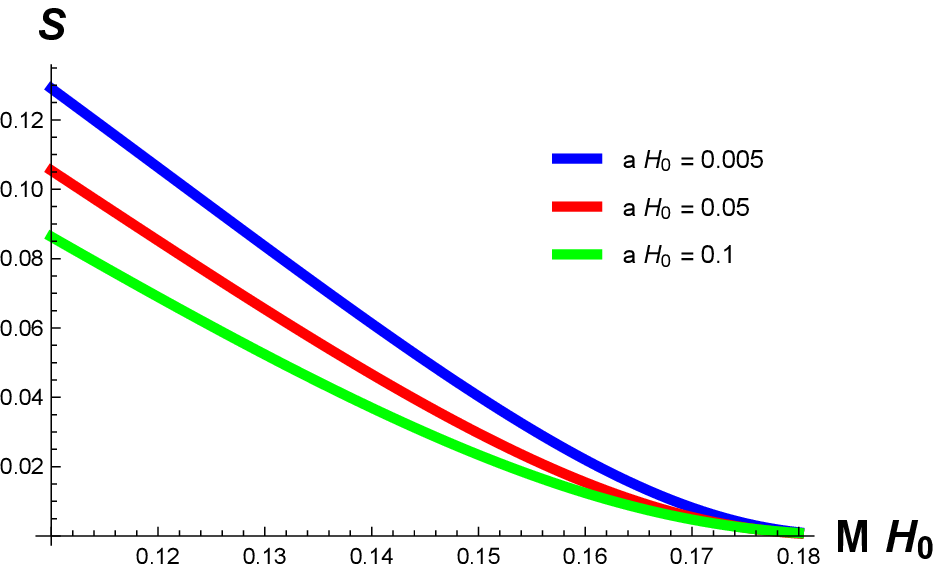}
  \caption{$ \frac{p}{H_0} = 0.5, m = -1/2$}
\end{subfigure}%
\begin{subfigure}{.45\textwidth}
 \centering
  \includegraphics[width=1\linewidth]{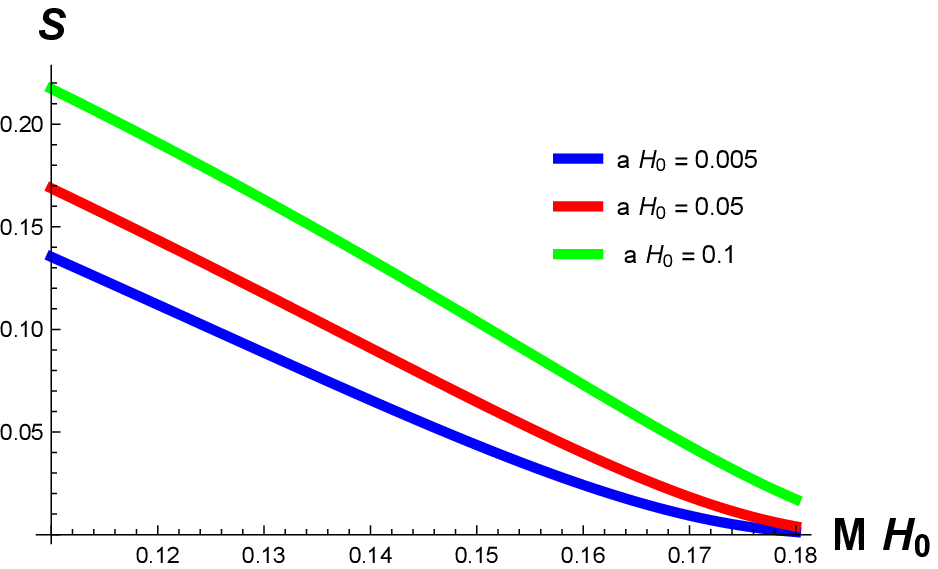}
  \caption{$ \frac{p}{H_0} = 0.5, m = 1/2$}
\end{subfigure}
\caption{The variation of the entanglement entropy, \ref{ee-kds}, with respect to the dimensionless mass parameter, $MH_0$.}
\label{fig4}
\end{figure}

We  now wish to investigate numerically the variation of $S(p,m)$ in \ref{ee-kds} numerically, with respect to $p$, $m$, and
as well as the spacetime parameters. If we assume a Kerr-de Sitter black hole spacetime, we have the bounds on the dimensionless parameters, $MH_0 \lesssim 0.2435$ and $a/M \lesssim 1$, e.g.~\cite{Bhattacharya:2017scw}. \ref{fig2} shows the monotonic decrease of $S(p,m)$ with respect to the increase of the dimensionless energy, $p/H_0$. \ref{fig3} and \ref{fig4} show the variation of $S(p,m)$ respectively with respect to the dimensionless rotation and mass parameters. In \ref{fig3}, the distinction of the positive and negative $m$-values are manifest, follows from the increase of both $\Omega_C$ and $\kappa_C$ with the increase of $aH_0$. This distinction is a quantum analogue of the distinction of the classical pro- and retrograde orbits~\cite{Chandrasekhar}. The decrease of $S(p,m)$ with respect to increasing $MH_0$ correspond to the fact that  $\kappa_C$ decreases  whereas $\Omega_C$ increases with the increase of the same.  

Finally, we  note that even though for the black hole event horizon \ref{ee-kds} is formally similar, 
its horizon parameters, $\kappa_H$, $\Omega_H$ show some qualitatively different variations with respect to $aH_0$ and $MH_0$.  Thus in this case, while \ref{fig2} will remain similar, {\it a priori} one might expect \ref{fig3} and \ref{fig4} will show  different  behaviour. However, this is not the case, as can be checked numerically. In fact the combination $(p-m\Omega_H)/\kappa_H$ appearing in the expression of the entanglement entropy makes it behave qualitatively exactly similarly as of \ref{fig3} and \ref{fig4}, for the black hole event horizon as well.  \\

The above result can be  extended to more general stationary axisymmetric spacetimes,
for example, the general  Plebanski-Demianski-de Sitter class~\cite{Griffiths:2005qp},
\begin{eqnarray}
ds^2=\frac{1}{\Omega^2}\left[-\frac{\Delta_r}{\rho^2}\left(dt-\left(a\sin^2\theta+4l\sin^2\frac{\theta}{2}\right)d\phi \right)^2+\frac{\rho^2}{\Delta_r}dr^2  + \frac{P}{\rho^2} \left(adt-\left(r^2 +(a+l)^2 \right) d\phi \right)^2      +\frac{\rho^2}{P}\sin^2\theta d\theta^2 \right]
\label{es18}
\end{eqnarray} 
where
\begin{eqnarray*}
\Omega&=&1-\frac{{\alpha}}{\omega}\left( l+a\cos\theta\right)r, \quad \rho^2=r^2+\left( l+a\cos\theta\right)^2, \quad
P=\sin^2\theta \left(1-a_3\cos\theta-a_4\cos^2\theta\right)\nonumber\\
\Delta_r&=&\left(\omega^2 k+q^2+q_m^2\right)-2Mr+\epsilon r^2 -\frac{2{\alpha} n}{\omega}r^3-\left({\alpha}^2 k+ H_0^2\right)r^4,
\label{es19}
\end{eqnarray*} 
The parameters ${\alpha}$, $\omega$, $q$, $q_m$, $\epsilon$ and 
$k$ are independent, and $a_3$ and $a_4$ are determined from them via a couple of constraints. Physical meaning to these parameters could be asserted for only certain special subclasses of \ref{es18}. 
For example for ${\alpha}=0$, the above metric reduces to the Kerr-Newman-NUT-de Sitter solution~\cite{Griffiths:2005qp},
\begin{eqnarray}
ds^2=-\frac{\Delta_r}{\rho^2}\left[dt-(a\sin^2\theta+4l \sin^2\frac{\theta}{2})d\phi\right]^2+\frac{\rho^2}{\Delta_r}dr^2
+\frac{P}{\rho^2}\left[adt-(r^2+(a+l)^2)d\phi\right]^2+\frac{\rho^2}{P}\sin^2\theta d\theta^2,
\label{es20}
\end{eqnarray} 
where
\begin{eqnarray*}
\rho^2&=&r^2+\left( l+a\cos\theta\right)^2, \quad
P=\sin^2\theta \left(1+ {4 a l H_0^2 \cos\theta}+ {H_0^2 a^2 \cos^2\theta}\right)\nonumber\\
\Delta_r&=&\left(a^2-l^2+q^2+q_m^2\right)-2Mr+r^2 -3H_0^2\left((a^2-l^2)l^2+\left(\frac{a^2}{3}+2l^2\right)r^2+ \frac{r^4}{3}\right),
\label{es21}
\end{eqnarray*} 
where $q$ and $q_m$ are respectively the electric and magnetic charge and $l$ is the NUT parameter.
We shall consider below the most general form of \ref{es18}, assuming implicitly it indeed
represents a well behaved spacetime possessing a cosmological event horizon horizon at $\Delta_r=0$. We shall not be concerned with the explicit parameter values corresponding to this assertion.

The  choice of the hypersurface orthogonal timelike vector field $\chi^{\mu}$ and the orthogonal basis for this case is exactly formally similar to the Kerr-de Sitter spacetime described above. Likewise, the R-L entanglement entropy turns out to be formally similar to \ref{ee-kds} in this case, with the horizon parameters
\begin{eqnarray*}
\Omega_C = \frac{a}{r^2_{C}+(a+l)^2} \qquad -\kappa_{C}= \frac{\Delta_r'}{2(r^2+a^2)}\bigg\vert_{r= r_C}
\label{es22}
\end{eqnarray*} 
%

\section{Conclusions} \label{concl}
In this work we have addressed the issue of the quantum entanglement for the Dirac fermions between the causally disconnected R-L regions of a static de Sitter spacetime. We have discussed the case of the simple de Sitter spacetime using the closed form mode functions. The behaviour of the fermionic  entanglement entropy, like that of a scalar field~\cite{Higuchi:2018tuk}, was shown to be similar to that of the Rindler spacetime~\cite{Alsing:2006cj, Mann:2009}. We further extended our result to the case of general static and spherically symmetric spacetime and as well as to the stationary axisymmetric spacetimes. For the Kerr-de Sitter spacetime in particular, we have numerically investigated the variation of the entanglement entropy with respect to the energy and angular momentum eigenvalues as well as the (dimensionless) spacetime parameters. The entanglement entropy for the non-extremal black hole horizon also shows similar variations with respect to these parameters.

 An interesting direction in which this work can be extended seems to be the consideration of charged fields in the presence of  background electromagnetic fields associated with a Killing horizon. We hope to come back to this issue in the near future.

\bigskip
\section*{Acknowledgement}
SG would like to acknowledge I.~Cotaescu and A.~Abrikosov~Jr. for helpful discussions. SB's research is partially supported by the ISIRD grant 9-289/2017/IITRPR/704. SC is partially supported by the ISIRD grant 9-252/2016/IITRPR/708.

\end{document}